\documentstyle[11pt,newpasp,epsf]{article}

\begin{document}

\title{The APM/Matched-Filter Cluster Catalog}
\author{Wataru Kawasaki}
\affil{Department of Astronomy, the University of Tokyo}

\begin{abstract}
A catalog of nearby clusters in the 5800 deg$^2$ area in the southern 
Galactic cap is constructed by applying a matched-filter cluster-finding 
algorithm to the sample of 3.3 million galaxies from the APM Galaxy 
Survey. 
I have preliminarily detected more than 4000 cluster candidates with 
estimated redshift of less than 0.2 and with richness similar to those 
of ACO clusters. Generally, a good correspondence is found between the 
nearest cluster candidates in our catalog and the ACO clusters which 
have measured redshift. While the ACO catalog becomes incomplete at 
$z>0.08$, the completeness limit of our cluster catalog reaches $z=0.15$. 
\end{abstract}


\section{Introduction}
Catalogs of galaxy clusters have provided useful data for numbers of 
works in the field of observational cosmology. 
The popular catalogs of nearby clusters include those by Abell(1958), 
Zwicky et al. (1961-68), Abell, Corwin, and Olowin (1989, hereafter 
ACO), Lumsden et al. (1992), Dalton et al. (1994). 
However, these catalogs are claimed to suffer from incompleteness 
and contamination because clusters in these catalogs are identified 
basically by finding local peaks of surface number density of galaxies 
in the sky. 
A new cluster-finding algorithm, matched-filter, was proposed by 
Postman et al. (1996) and it is becoming a standard way for finding, 
especially, distant clusters. 
This time I have applied matched-filter to the 3.3 million galaxy 
sample from the APM Galaxy Survey (Maddox et al. 1990a, 1990b) and 
constructed a new catalog of nearby clusters covering about 5800 
deg$^2$ in the southern Galactic cap. 

\section{Cluster Finder}
In this work, a matched-filter algorithm by Kawasaki et al. (1998) 
is used. 
Simultaneously using two-dimensional positions and apparent 
magnitudes of galaxies, matched-filter has several advantages 
over the classical cluster-finding algorithms such as count-in-cells 
which use only galaxy positions: (i) much poorer clusters which were 
undetectable with classical techniques can be found, (ii) spurious 
detection rate of non-physical clusters is greatly suppressed, and 
(iii) rough estimates of redshift and richness of clusters are 
obtained as byproducts of detection (for details, see Kawasaki et al. 
1998). 

\begin{figure}
\vspace*{8cm}
\includegraphics{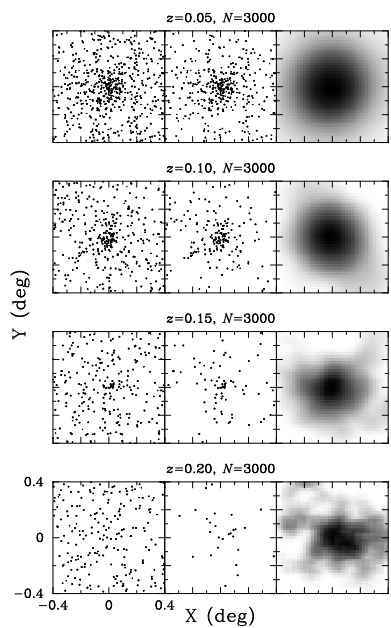}
\caption{Four artificial Coma-like clusters at different redshifts. 
Left, middle, and right columns show the whole (cluster+field) galaxy 
distributions, galaxy distributions for cluster members only, and 
`richness maps', output of the matched-filter, respectively. }
\end{figure}

\begin{figure}
\vspace*{8cm}
\includegraphics{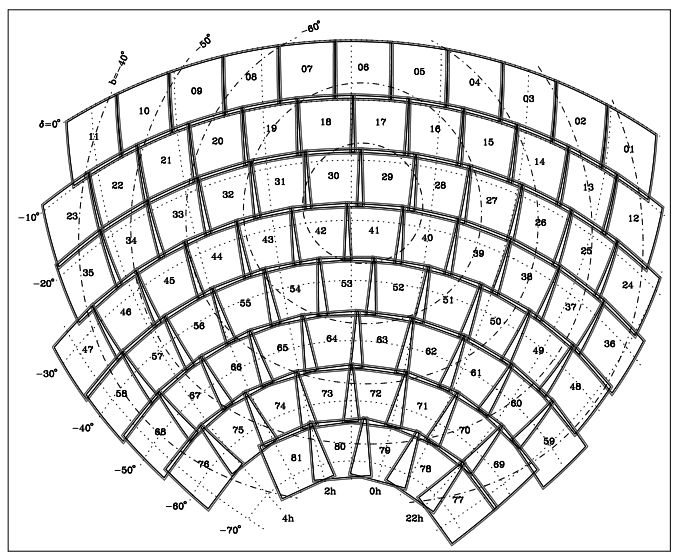}
\caption{Location of the 81 10$\deg \times$10$\deg$ subregions in an 
equal area projection. The thin solid lines show the edges of the 
subregions. For each subregions the inner 9$\fdg$44$\times$9$\fdg$44 
area in which clusters do not suffer from vignetting effect are enclosed 
with thick solid lines. Serial numbers are given for each subregion. }
\end{figure}

\begin{figure}
\vspace*{8.5cm}
\includegraphics{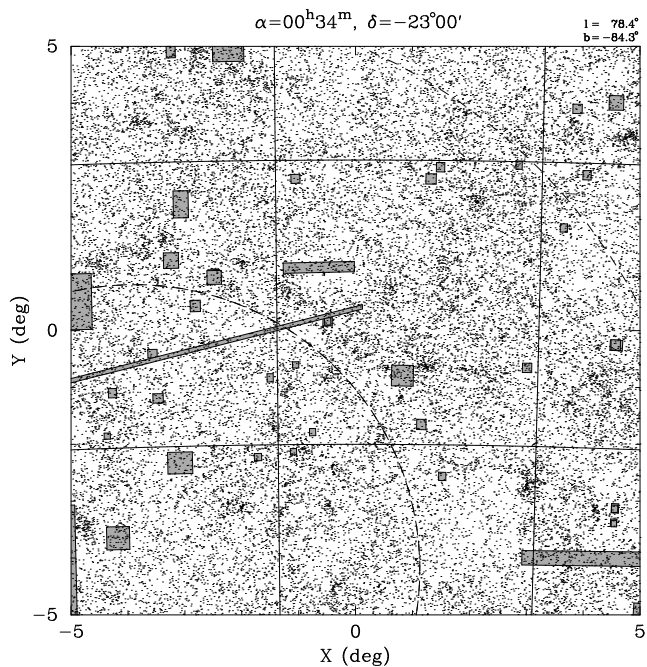}
\caption{The distribution of the APM galaxies with extinction-corrected 
magnitudes of brighter than $b_j = 20.0$ in the 10$\deg \times$10$\deg$ 
area near the South Galactic Pole (subregion \# 29). }
\end{figure}

\begin{figure}
\vspace*{8.5cm}
\includegraphics{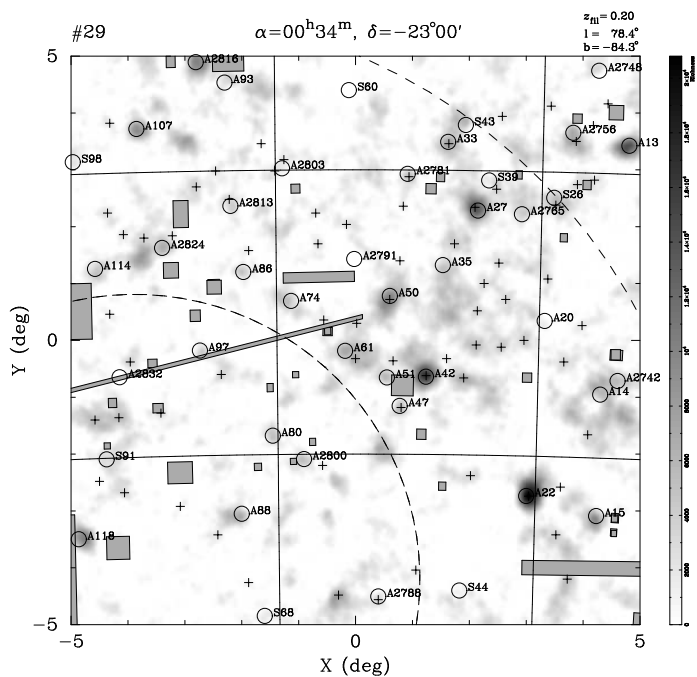}
\caption{A `richness map' for the area shown in Figure 3. The pluses are 
the cluster candidates with $z_{\rm est} \leq 0.20$ and $N_{\rm est} 
\geq 1000$. Abell/ACO clusters are shown with open circles. }
\end{figure}

\begin{figure}
\vspace*{8cm}
\includegraphics{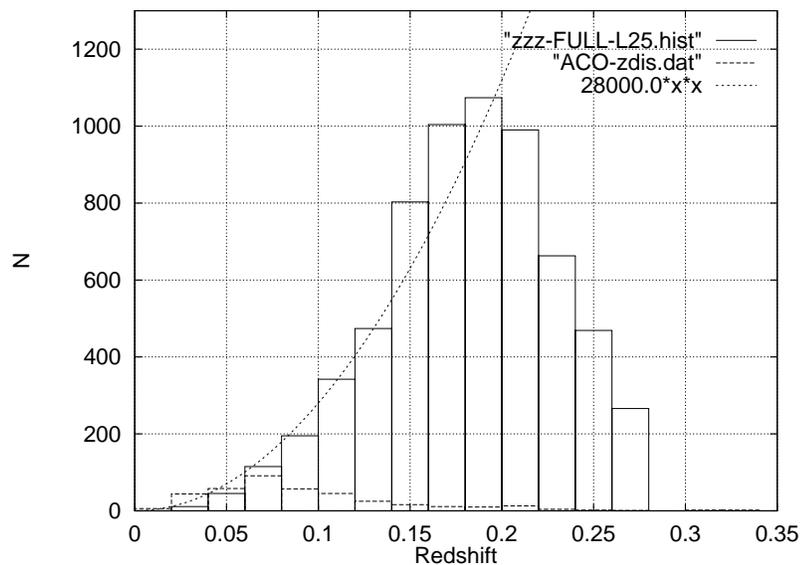}
\caption{The redshift distribution of the cluster candidates is 
shown with the solid line, while that of Abell/ACO clusters with 
spectroscopically measured redshift is shown with the ling-dashed 
line. The short-dashed line proportional to $z^2$, meaning constant 
number density, is arbitrarily fit to the left-side of the histograms. }
\end{figure}

\begin{figure}
\vspace*{9cm}
\includegraphics{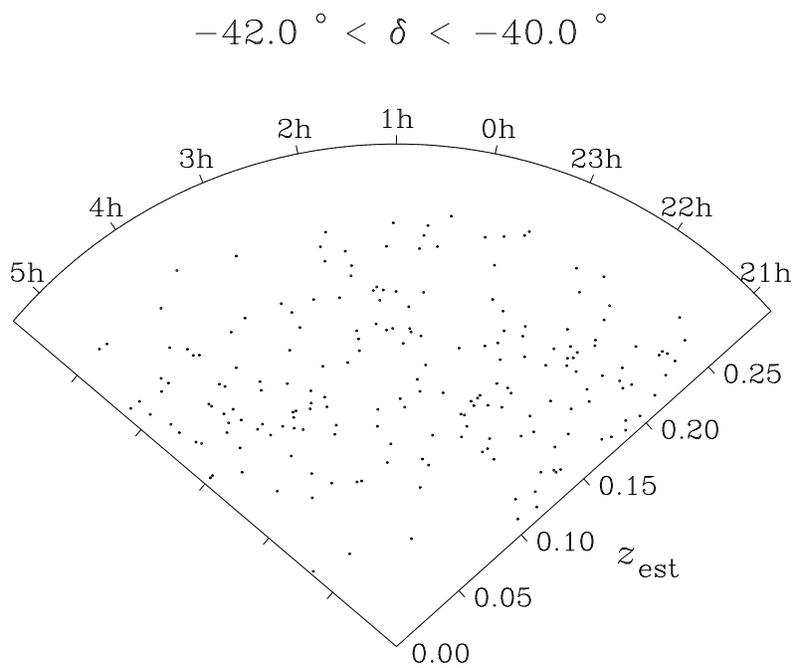}
\caption{A wedge diagram for the cluster candidates. }
\end{figure}

Figure 1 shows a demonstration of matched-filter's ability of 
finding distant/poor clusters which count-in-cells can not find. 

\section{Catalog Construction}
As the unit of processing the whole data, 81 subregions with 
10$\deg \times$10$\deg$ area are defined as shown in Figure 2. 
The galaxy distribution in one of the subregion is shown in 
Figure 3, and a `richness map', the output of matched-filter, is 
shown in Figure 4. The cluster candidates appear as peaks in the 
gray scale. Here cluster candidates with estimated richness 
$N_{\rm est} \geq 1000$ (similar to Abell Richness Class 0) are 
compiled to the preliminary catalog. they are shown with the plus 
marks in Figure 4. 

\section{Properties of the Preliminary Catalog}
For nearby apparent clusters, Figure 4 shows that there is 
generally a good correspondence between our sample and Abell/ACO 
clusters. However, it is also shown that some apparent candidates 
are missed in ACO catalog. ACO clusters which are not cataloged 
in our sample are less rich ($N_{\rm est} < 1000$) ones. 
Our cluster sample seems to be complete up to $z \sim 0.15$, twice 
deeper than the completeness limit for ACO clusters with measured 
redshifts. For $z > 0.15$, there may exist a significant fraction 
of contamination. Incidentally noting, Figure 6 seems to show the 
non-uniformity of cluster distribution on 100Mpc scale. 

\acknowledgments
I am grateful to Drs. Steve J. Maddox and Mike J. Irwin for 
providing me with the APM galaxy catalog, the entire data 
resource for this work.

\end{document}